\documentclass[usegraphicx]{mn2e}

\usepackage{times, amssymb}

\title[Optical and near-infrared spectroscopy of V4641~Sgr]{New clues on outburst 
mechanisms and improved spectroscopic elements of the black-hole binary V4641~Sagittarii
\thanks{Based on data obtained at the Mount Stromlo and Siding Spring Observatories}}
\author[C. Lindstr\o m et al.]{C. Lindstr\o m$^1$, J. Griffin$^1$,
L. L. Kiss$^1$\thanks {On leave from University of Szeged,
Hungary}\thanks{E-mail: \tt l.kiss@physics.usyd.edu.au}, M. Uemura$^{2,3}$, 
A. Derekas$^{1,4}$, Sz. M\'{e}sz\'{a}ros$^5$, \newauthor P. Sz\'{e}kely$^6$\\
\\
$^1$School of Physics A28, University of Sydney, NSW 2006, Australia\\
$^2$Departamento de F\'{i}sica, Facultad de Cienc\'{i}as F\'{i}sicas y Matem\'{a}ticas,
Universidad de Concepci\'{o}n, Casilla 160-C, Concepci\'{o}n, Chile\\
$^3$Yukawa Institute for Theoretical Physics, Kyoto University, Kyoto 606-8502, Japan\\
$^4$School of Physics, Department of Astrophysics and Optics, University of
New South Wales, NSW 2052, Australia\\
$^5$Department of Optics and Quantumelectronics, University of Szeged, Hungary\\
$^6$Department of Experimental Physics and Astronomical Observatory, University of
Szeged, Hungary}

\begin{document}

\date{Accepted ... Received ..; in original form ..}


\maketitle

\begin{abstract}

We present spectroscopic observations of the black-hole binary V4641~Sagittarii,
obtained between 4th July 2004 and 28th March 2005, which cover the minor outburst of
the star in early July 2004 and quiescence variations on 19 nights scattered over six
months. During the outburst, the star peaked approximately 3 magnitudes brighter
than usual, and our spectra were dominated by broad hydrogen, helium and iron  emission
lines. The very first spectra showed P~Cygni profiles, which disappeared  within a few
hours, indicating rapid changes in matter ejection. The H$\alpha$ line had multiple
components, one being a broad blue-shifted wing exceeding 5000 km/s. During a
simultaneously observed 10-min photometric flare-up, the equivalent width of the
H$\alpha$ line temporarily decreased, implying that it was a flare of the continuum. The
overall spectral appearance was similar to that observed in the 1999 September active
phase, which suggests that similar mass-ejection processes were associated with both
eruptions. In quiescence, the spectra were those of the early-type secondary star
showing its orbital motion around the primary. By measuring cross-correlation radial
velocities, we give an improved set of spectroscopic elements. Whereas we measure the 
same velocity amplitude ($K_2=211.3\pm1.0$ km~s$^{-1}$), within errors,  as
Orosz et al. (2001), our centre-of-mass velocity ($\gamma=72.7\pm3.3$ km~s$^{-1}$) 
differs significantly from the previously published value ($107.4\pm2.9$ km~s$^{-1}$).
However, we find evidence that the difference is caused by a systematic error 
in data reduction in the previous study, rather than by gravitational effects of 
an invisible third component.

\end{abstract}

\begin{keywords}
X-rays: binaries -- black hole physics -- stars: individual: V4641~Sgr
\end{keywords}

\section{Introduction}

The X-ray transient V4641~Sgr (SAX~J1819.3--2525) is a unique  black-hole binary, which
has attracted considerable interest since its major outburst in September 1999, where it
peaked
at $V\sim8.8$  (Stubbings 1999). Even in quiescence, the star is fairly bright at
$V\approx13.7$ mag. It has an orbital period of 2.817 days, and is located at a distance of
$\sim7-12$ kpc toward the galactic bulge (Orosz et al. 2001). The total mass of the
system lies between 14--20 M$_\odot$ with a black hole primary of $\sim9-12$ M$_\odot$
(Orosz et al. 2001). Given its moderate inclination (60--70$^\circ$, Orosz et al.
2001), the system does not show eclipses and the quiescence light curve is dominated by
ellipsoidal variability. The secondary was classified as a B9III star, being one of the
most luminous secondary stars known in dynamically established black-hole binaries (Orosz
et al. 2001). The large eruption in 1999 was associated with superluminal jet ejection
and was followed by a series of smaller outbursts in the last few years. These involved 
violent flashes and rapid variability, making this relatively bright system a key  target
for understanding the processes associated with black hole binaries.

The activity of V4641~Sgr has been variable since the 1999 September 
outburst. As a faint X-ray source it was discovered independently by RXTE  (Markwardt
et al. 1999) and BeppoSax (in't Zand et al. 1999) in 1999 February. The 1999 September
event, peaking at 12.2 Crab in the 2--12 keV band, was extremely rapid in its rise and
decay across radio,  optical and X-ray wavelengths. In X-rays, a brief but dramatic
eruption occured on Sept. 15-16, 1999, showing remarkably rapid fluctuations (Wijnands
\& van der Klis 2000). The  optical decay lasted about two days (e.g. Chaty et al. 2003),
while the radio transient was detected for a further three weeks. The measurable proper
motion  of the resolved radio structures indicated superluminal jet ejection (Hjellming
et al. 2000). Subsequent quiescence observations led Orosz et al. (2001) to measuring
the most accurate system parameters so far, which revealed the massive stellar black
hole primary and the nature of the early-type secondary star.     

The second major outburst occurred in 2002 May, when unprecedented rapid optical 
fluctuations were detected (Uemura et al. 2002). The short time-scale of the
large-amplitude variations indicated that the inner region of the accretion disk made
a significant contribution to the optical flux, implying strong non-thermal emission
in the optical (Uemura et al. 2002, 2004b).  Another outburst in 2003 August was very
similar to the one in the previous year, leading Uemura et al. (2004a) to the
conclusion that the two eruptions had the same nature. They also suggested the
possibility of recurrent outbursts on a  time-scale of 1--2 years.

This estimate of the time-scale was supported by R.Stubbings's visual report of
a fourth major outburst on 2004 July 4.368 UT, where the star peaked at around
$V\approx11.1$ mag  (Uemura et al. 2005).  Coincidentally, the star was
detected with the RXTE at 8.2 mCrab flux (2-10 keV) on July 3.496, roughly
eight times fainter than during the  outburst in 2003 (Swank 2004) and three
orders of magnitude dimmer than in the large  1999 September event. VLA
observations clearly detected the outburst in radio, as a 12.5 mJy source at
1.425 GHz, with a steep spectrum (Rupen et al. 2004).  A day later, V4641~Sgr
was already close to its quiescence level (Revnivtsev et al. 2004), although
very rapid, post-outburst flares of 1 mag were seen as late as  17 days after
the main outburst (Bikmaev et al. 2004, Uemura et al. 2005). Most recently, a
new active phase occured in late June, 2005, when V4641~Sgr was detected  in a
very similar outburst like those in 2002, 2003 and 2004 (Swank et al. 2005,
Khamitov et al. 2005, Buxton et al. 2005).

We started spectroscopic monitoring of the system on 2004 July 4.48, roughly two hours
after the 2004 outburst was reported via electronic mailing lists, with a few optical 
and near-infrared spectra (an early report was given by Kiss \& M\'esz\'aros 2004). In all
follow-up observations between 2004 September and 2005 March were taken in the
quiescence of V4641~Sgr and they allowed us to measure an improved set of spectroscopic
elements. 

Some of the important questions we wish to address in this paper are as
follows. What is the driving mechanism behind these recurrent outbursts? What sort of
processes accompany an eruption? How much are the 2002--2004 outbursts different from
the major one in 1999 September? Which system parameters can be better
constrained using our dataset? We describe the  observations in Sect.\ 2; Sect.\ 3
discusses the outburst spectra, while quiescence  data are presented in Sect.\ 4.
Discussion of the above questions can be found in Sect.\ 5 and the paper concludes in Sect.\ 6.

\section{Observations and data reductions}

\begin{table*}  
\begin{centering}
\caption{Observation log. `HJD' refer to the range in heliocentric Julian Dates, marked
by the first and the last spectrum on any given night.} 
\label{log}  
\begin{tabular}[b]{|l|l|c|c|c|c|c|c|}  
\hline 
Date & HJD (2,453,000+) & \multicolumn{2}{|c|}{Red arm} & \multicolumn{2}{|c|}{Blue arm} \\

 && Wavelength & No. of & Wavelength & No. of \\
 && range (\AA) & spectra & range (\AA) & spectra \\
\hline 

\textit{Outburst} \\
2004 July 4 & 190.9870--191.2176 & 8043-8972 & 4 & & \\
            & 191.2271--191.2652 & 6047--7005 & 10 & & \\
	    & 190.9870--191.2629 &   & & 4940--5383 & 10\\
  
\textit{Quiescence} \\
2004 Sept 25 & 274.013--274.066 & 5786--6748 & 4 & 4510--4973 & 4 \\
2004 Sept 26 & 274.882--275.038 & 5786--6748 & 8 & 4510--4973 & 8 \\
2004 Sept 27 & 275.878--276.027 & 5786--6748 & 2 & 4510--4973 & 2 \\
2004 Sept 28 & 276.878--277.026 & 5786--6748 & 7 & 4510--4973 & 7 \\
2004 Oct 25 &303.894& 5794--6756 & 1 & 4785--5233 & 1 \\
2004 Oct 26 &304.917--304.932 & 5794--6756 & 2 & 4785--5233 & 2 \\
2004 Oct 27 & 305.899--305.916 & 5794--6756 & 2 & 4785--5233 & 2 \\
2004 Oct 28 & 306.894--306.908 & 5794--6756 & 2 & 4785--5233 & 2 \\
2004 Oct 29 & 307.889--307.903 & 5794--6756 & 2 & 4785--5233 & 2 \\
2004 Oct 30 & 308.913--308.927 & 5794--6756 & 2 & 4785--5233 & 2 \\
2004 Oct 31 & 309.905 & 5794--6756 & 1 & 4785--5233 & 1 \\
2004 Nov 1 & 310.899 & 5794--6756 & 1 & 4785--5233 & 1 \\
2004 Nov 2 & 311.915 & 5794--6756 & 1 & 4785--5233 & 1 \\
2005 Mar 21 & 451.232--451.247 & 5794--6756 & 2 &          &  \\
2005 Mar 24 & 454.245--454.260 & 5794--6756 & 2 &          &  \\
2005 Mar 25 & 455.272--455.287 & 5794--6756 & 2 &          &  \\
2005 Mar 26 & 456.260--456.275 & 5794--6756 & 2 &          & \\
2005 Mar 27 & 457.195 & 5794--6756 & 1 &          & \\
2005 Mar 28 & 458.199--458.213 & 5794--6756 & 2 &           & \\
\hline 
\end{tabular}
\end{centering}
\end{table*}

The observations were carried out with the 2.3m telescope at the Siding Spring
Observatory, Australia. Four sets of observations were obtained: 2004 July 4, during the
outburst; four nights in late 2004 September, nine nights in 2004 October/November and
six nights in 2005 March, all in quiescence (see Table \ref{log} for
details). All spectra were taken with the Double Beam Spectrograph using 1200 mm$^{-1}$
gratings in both arms (except in 2005 March, when a CCD failure
prevented the use of the blue arm). The exposure time ranged between 240--900 s for the
outburst spectra and was 1200 s in quiescence. The dispersion was 0.55 \AA\ px$^{-1}$,
leading to a nominal resolution of about 1 \AA. In addition to the V4641~Sgr spectra, we
regularly observed a telluric standard (HD~177724) and several radial velocity standard
stars (HD~187691, $\beta$~Vir, HR~3383). The latter were used to measure
cross-correlation radial velocities in quiescence.

All spectra were reduced with standard tasks in  IRAF\footnote{IRAF is distributed by
the National Optical Astronomy Observatories, which are operated by the Association of
Universities for Research in Astronomy, Inc., under cooperative agreement with the
National Science Foundation.}. Reduction consisted of bias and flat field corrections,
aperture extraction,  wavelength calibration and continuum normalization. We did not
attempt flux  calibration because the conditions were generally non-photometric. We
checked the consistency of wavelength calibrations via the constant positions of strong 
telluric features, which proved the stability of the system. Radial velocities were
determined with the task {\it fxcor}, including barycentric corrections.
Different velocity standards have shown that our absolute velocity frame was stable
to within $\pm$2--3 km~s$^{-1}$.

\section{Spectra of the 2004 July outburst}

About two hours after the first visual report arrived, we took three spectra in both 
the blue and red arms (between July 4.48-4.50 UT). The latter was observing in the
near-infrared, showing spectra dominated by the Paschen series. Five hours later, we
went back to the star to take another near-infrared spectrum, after which we switched to
the H$\alpha$ region. Thereafter we took spectra in both arms continuously until the
object was too low over the horizon. We obtained a total of 24 spectra, which showed
remarkable changes with time.  Orbital phases were assigned using the following
ephemeris for the shallower minimum of the ellipsoidal variability (Uemura et al., in
preparation; see also Sect.\ 4 for the spectroscopic ephemeris): 

\begin{equation}
{\rm HJD}_{\rm min}=2453154.67359 + 2.81728 \times E
\end{equation}

\noindent Using this ephemeris, the orbital phases covered by our data ranged between
0.89 and  0.99, where 1.00 corresponds to the secondary passing in front of the black
hole.    

We show a representative collection of the three spectral regions observed during 
outburst in Figs\ \ref{tp1}--\ref{tp2} and Fig.\ \ref{tp3}. Lines of hydrogen and helium
dominate the spectrum with weaker detections of Fe II, Mg II and Si II. Rapid variations
in line profiles between spectra indicate a high and unpredictable activity, 
whereas a large range of gas velocities is reflected in the broad emission 
profiles.

\subsection{Blue region}

\begin{figure}
\begin{center}  
\leavevmode  
  \includegraphics[width=80mm]{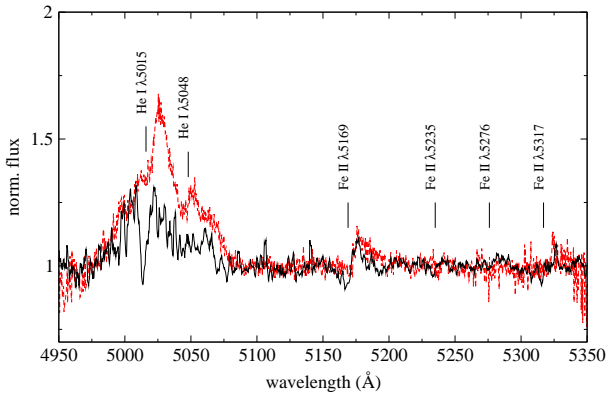}
\end{center}
\caption{The blue region at July 4.481 UT (black/solid line) and 6.4 hours later at 
July 4.748 UT (red/dashed line).}
\label{tp1}
\end{figure}

Here we identify the strong blend of the He~I~$\lambda$5015 and 
$\lambda$5048 lines and four Fe~II lines, of which the Fe~II~$\lambda$5169 is the
strongest (Fig.\ \ref{tp1}), while Fe~II~$\lambda$5235 and
Fe~II~$\lambda$5276 are marginally detected. Line profiles of He~I~$\lambda$5015 
exhibited a rapid change between the early and the late phase, separated by more than
five hours. During the early phase, it contained a red-shifted peak, a deep absorption and a
blue-shifted peak. The blue peak apparently has a narrow peak with a broad shoulder.
During the late phase, the line profile becomes a complex emission line with a strong
red peak, a blue component and a broad blue wing. The full width at zero intensity
(FWZI) reached about 5000 km~s$^{-1}$, although it is difficult to measure due to the
close blend.

Of the four Fe~II lines, the strongest Fe~II~$\lambda$5169 line showed a clear 
P~Cyg-like profile in the early phase and possibly in the late phase. Similar features
are also visible in the Fe~II~$\lambda$5317 line. The absorption components were
blue-shifted by 500--600 km~s$^{-1}$, relative to the emission peak. 

\subsection{H$\alpha$ region}

\begin{figure}
\begin{center}  
\leavevmode  
  \includegraphics[width=80mm]{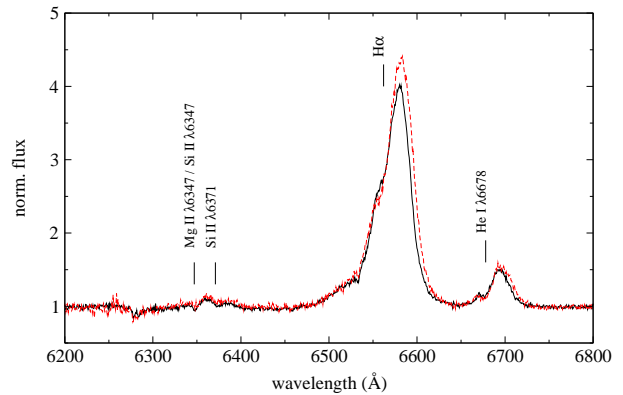}
\end{center}
\caption{The H$\alpha$ region at July 4.727 UT (black/solid line) and 
July 4.765 UT (red/dashed line).}
\label{tp2}
\end{figure}

\begin{figure}
\begin{center}  
\leavevmode  
  \includegraphics[width=80mm]{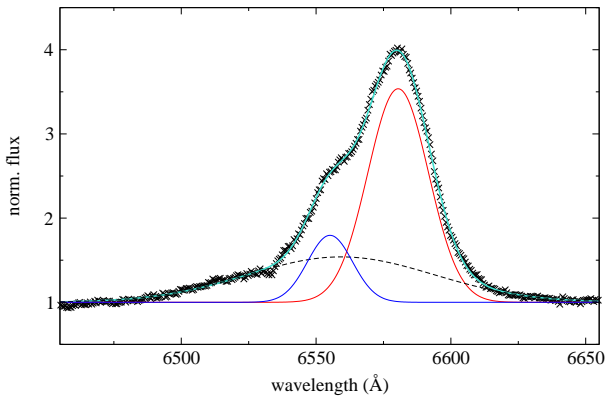}
\end{center}
\caption{Three-component Gaussian fit to the H$\alpha$ line.}
\label{fit}
\end{figure}

This region was dominated by a very broad and strong H$\alpha$ line, the
He~I~$\lambda$6678 line and weak Mg~II/Si~II emission features. The FWZI of the
H$\alpha$ line exceeded 5000 km~s$^{-1}$, with an equivalent width (EW) of about
130--140 \AA. The asymmetric line profile of the H$\alpha$ line can be explained by the
presence of a few (at least three) distinct components, namely a strong red-shifted
component, a weaker blue-shifted component and a broad wing. The He line had
approximately the same structure. Both profiles suggest the presence of a high-velocity
outflow component, which is  strikingly similar to what was reported in the 1999
September outburst (Orosz et al. 2001, Chaty et al. 2003).

\begin{figure}
\begin{center}  
\leavevmode  
  \includegraphics[width=85mm]{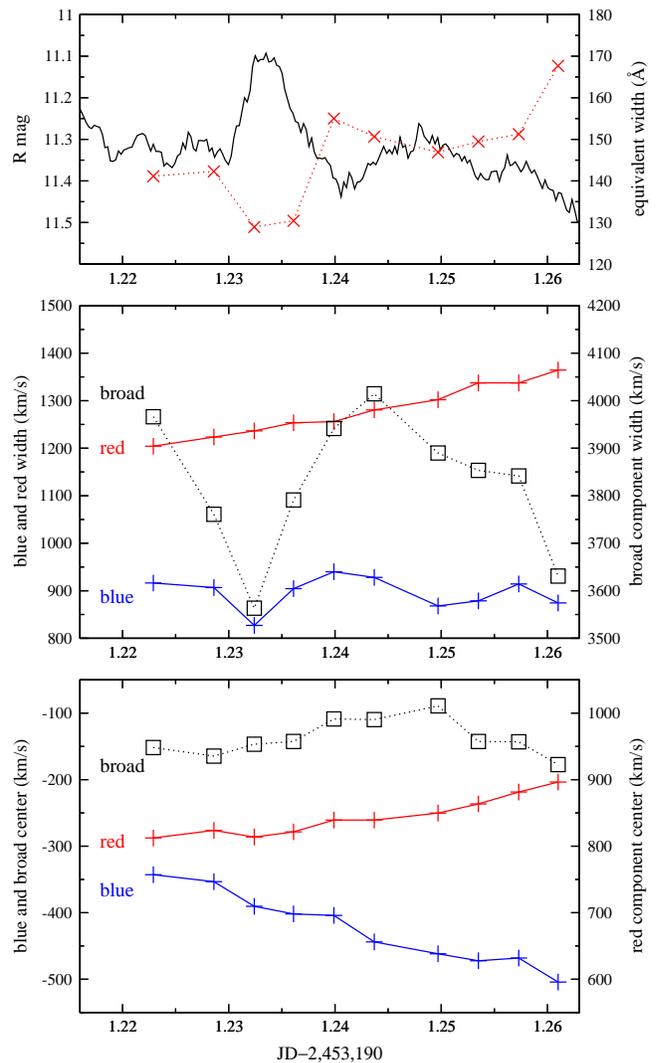}
\end{center}
\caption{Line profile changes correlated with light variability. {\it Upper panel:} 
equivalent width of the H$\alpha$ line (red crosses with dotted line) and 
observed apparent magnitudes (solid line). The typical
error of the equivalent width is a few \AA. {\it Middle panel:} Time variations of 
the FWHMs of the three Gaussian components, expressed in km~s$^{-1}$. Blue and red
crosses show the blue and red components, while open squares refer to the broad 
component.
{\it Bottom panel:} Time variations of the line centres of the Gaussian components.
Symbols as in the middle panel.
}
\label{flare}
\end{figure}

The roughly one hour of continuous observations in the H$\alpha$ region revealed further
interesting features. Firstly, there was a significant change
in the line profile over the one hour interval. Secondly, from Fig.\ \ref{tp2}, it is apparent that the red peak moved 
further redward, while the  blue component moved significantly blueward. The
width of the red component also increased with time, while very similar changes occurred
in the He~I~$\lambda$6678 line. To quantify these changes, we fitted multiple
Gaussian components to all 10 H$\alpha$ profiles. A sample fit is shown
in Fig.\ \ref{fit}. It turned out that the red peak moved by $\sim$80 km~s$^{-1}$
towards red, while the blue peak moved almost twice as much, $\sim$150 km~s$^{-1}$
towards blue. Their widths also changed from spectrum to spectrum, indicating significant
changes in line profiles.

In parallel with these H$\alpha$ spectra, one of us (M.U.) 
made simultaneous CCD photometric observations in Chile, using a
30 cm telescope at Universidad de Concepcion. These observations 
covered a 10-minute flare with 0.2 mag amplitude at JD=2,453,191.233, which just
coincided with taking two spectra in Australia. The one-hour interval of simultaneous 
photometry and spectroscopy allowed us to investigate unprecedented detail into the 
rapid variability of the system (the full photometric behaviour of the 2004 eruption is 
discussed in Uemura et al. 2005).  

We correlate the equivalent width of the H$\alpha$ line and parameters of the fitted
Gaussian components with the light variations in Fig.\ \ref{flare}. The top panel shows
that there was an inverse correlation between the system's apparent magnitude and the
H$\alpha$ equivalent width. This indicates that the flare-up was largely a flare of 
continuum. The middle panel in Fig.\ \ref{flare} suggests, however, that line profile
changes also accompany the flaring activity: of the three components, the blue and the
broad ones narrowed significantly in the flare maximum. The formal error of the
widths of the Gaussian fits is about 2-3\%, which was definitely exceeded by the
observed $\sim11$\% changes. It is very interesting that the width of the red component
was very stable apart from a gradual broadening with time, indicating that it 
was formed in a completely different part of the system that did not take part in the
flaring activity. Finally, the bottom panel in Fig.\ \ref{flare} shows the variations of
the Doppler-shifts of the components with time. The individual errors are between 
10--50 km~s$^{-1}$, so the observed changes are real. Interestingly, they do not
show any visible correlation with the flaring activity, suggesting the lack of strong
asymmetries in line profile variations.

\subsection{Paschen lines}

\begin{figure}
\begin{center}  
\leavevmode  
  \includegraphics[width=80mm]{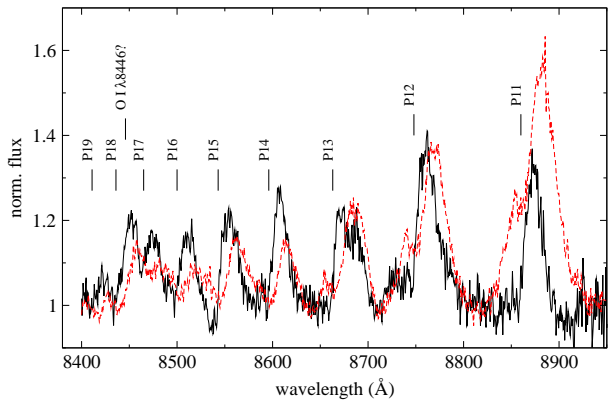}
\end{center}
\caption{Lines of the Paschen series at July 4.487 UT (black/solid line) and July 4.718
UT (red/dashed line). Numbers indicate the order within the series.}
\label{tp3}
\end{figure}

In the near-infrared spectra we identify higher order lines of the Paschen series, all
in emission, the He~II~$\lambda$8237 line with a P~Cygni profile and possibly some
emission of the O~I~$\lambda$8446 line. Due to the broadness of the features, there might be additional contamination from other species. For example, the P16, 15 and 13 lines
could be contaminated by calcium emmissions at 8498, 8542 and 8652 \AA,
respectively.We show a zoomed plot of the Paschen lines in Fig.\ \ref{tp3}, where the 
dashed-line spectrum was taken 5 hours later.

The Paschen lines changed significantly over the 5 hours between July 4.487
and 4.718 UT. These changes were similar both to those of the He~I lines in the blue region and the He~I~$\lambda$6678 line in the H$\alpha$ region, showing stronger
red-shifted and blue-shifted emission components. The FWZI increased from 1000--1500
km~s$^{-1}$ up to 3000 km~s$^{-1}$, which is also similar to that of the 
He~I~$\lambda$6678 line. Finally, whereas the earlier spectrum showed comparable line
strengths for the Paschen lines, this changed dramatically later, with the lower
members of the series (P11 and P12) becoming much stronger than the other lines. This is indicative
of switching from optically thick to optically thin emission, where the latter  is
characterized by a regular flux increase toward the lower member of the series (see
Storey \& Hummer 1995 for theory and Clark et al. 1998 for an application in a Be/X-ray
binary). We return to the implications of these spectra in Sect.\ 5.

\section{Radial velocities in quiescence}

We continued regular observations of V4641~Sgr after it returned to quiescence. Our
monitoring  has resulted in further 46 spectra, taken between 2004 Sept. 25 and 2005
March 28. None of the spectra showed any sign of variability or spectral line distortion
that might have occurred due to some accretion activity (Goranskij et al. 2003), with the
only detectable change due to the orbital motion of the second component. Given that the
one published radial velocity curve (Orosz et al. 2001) had quite large uncertainties
(see fig.\ 2 in Orosz et al. 2001), we decided to re-determine spectroscopic elements of
the system. 

\begin{figure}
\begin{center}  
\leavevmode  
  \includegraphics[width=80mm]{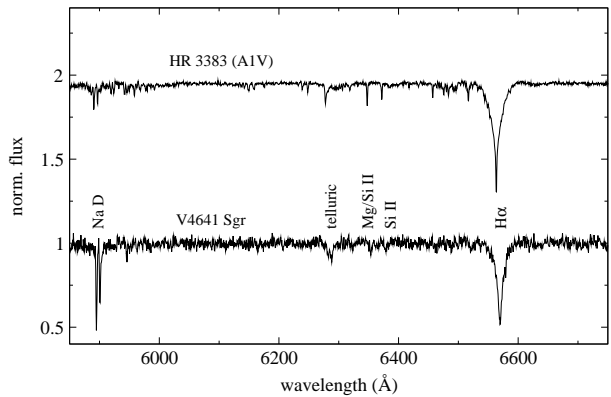}
\caption{A sample DBS-red spectrum of HR~3383 (upper plot) and V4641~Sgr in quiescence
(lower plot). }
\label{quiesc}
\end{center}
\end{figure}

A sample quiescence spectrum is shown in Fig.\ \ref{quiesc}. It is dominated by the
strong H$\alpha$ absorption line, which is characteristic for a late B/early A type star.
The strong Na~I~D doublet is of interstellar origin as its radial velocity  does not
change  with the orbital phase. Apart from these, only weaker features of Mg II and Si
II are easily identifiable, which agrees very well with the quiescence observations of
Orosz et al. (2001). As a comparison, we also plot a spectrum of HR~3383, an A1V star.
The close similarity of the spectra supports the spectral type determination by
Orosz et al. (2001).

\begin{figure}
\begin{center}  
\leavevmode  
  \includegraphics[width=80mm]{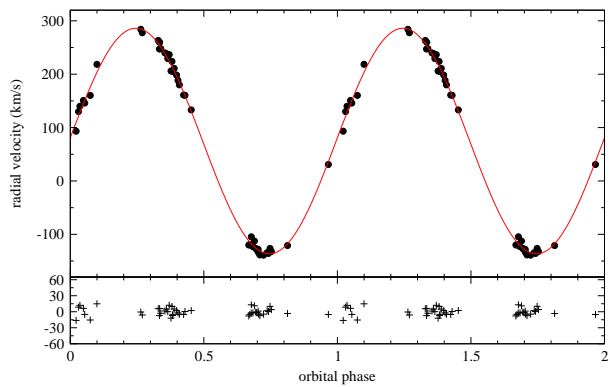}
\caption{New radial velocity data phased according to the ephemeris in Eq.\ (1). 
The rms of the sine-wave fit is 7 km~s$^{-1}$.}
\label{radvel}
\end{center}
\end{figure}

We decided to use the H$\alpha$ line for measuring radial velocities because: 1. the
H$\alpha$ region was observed on every night, providing a homogeneous  dataset; 2. this
region is where the B9III-type secondary (Orosz et al. 2001) and the A/F-type radial
velocity standards have the most similar spectra, minimizing the effects of spectral
mismatch on cross-correlation (Verschueren et al. 1999); 3. all the other stellar
features are almost lost in continuum scatter.  

We determined heliocentric radial velocities using 200 \AA\ of the spectra,  centred on
the H$\alpha$ line. As the primary standard we chose HD~187691, an  F8V star, while
$\beta$~Vir (F9V) and HR~3383 (A1V) served as check stars for the absolute accuracy. We
preferred the F-type standard over HR~3383 because of its  narrower H$\alpha$ line,
producing a slightly narrower cross-correlation peak.  A period search of the
radial velocities resulted in $P_{\rm spec}=2.817(2)$ d, which is in good 
agreement with the period determined by Orosz et al. (2001), and is practically
identical to the (more accurate) photometric period. For that reason, radial
velocities  were phased using the ephemeris in Eq.\ 1 and are plotted in Fig.\
\ref{radvel}. To estimate the measurement error, we took the residual mean scatter of
the sine-wave fit (solid line in Fig.\ \ref{radvel}) and the scatter of the interstellar
Na~I~D velocities, which were also determined with cross-correlation. Both estimates
gave consistent results ($\sigma_{V_{\rm r}}\approx7$ km~s$^{-1}$) and the symbol sizes
in Fig.\ \ref{radvel} reflect this uncertainty.

\begin{figure*}
\begin{center}  
\leavevmode  
  \includegraphics[width=160mm]{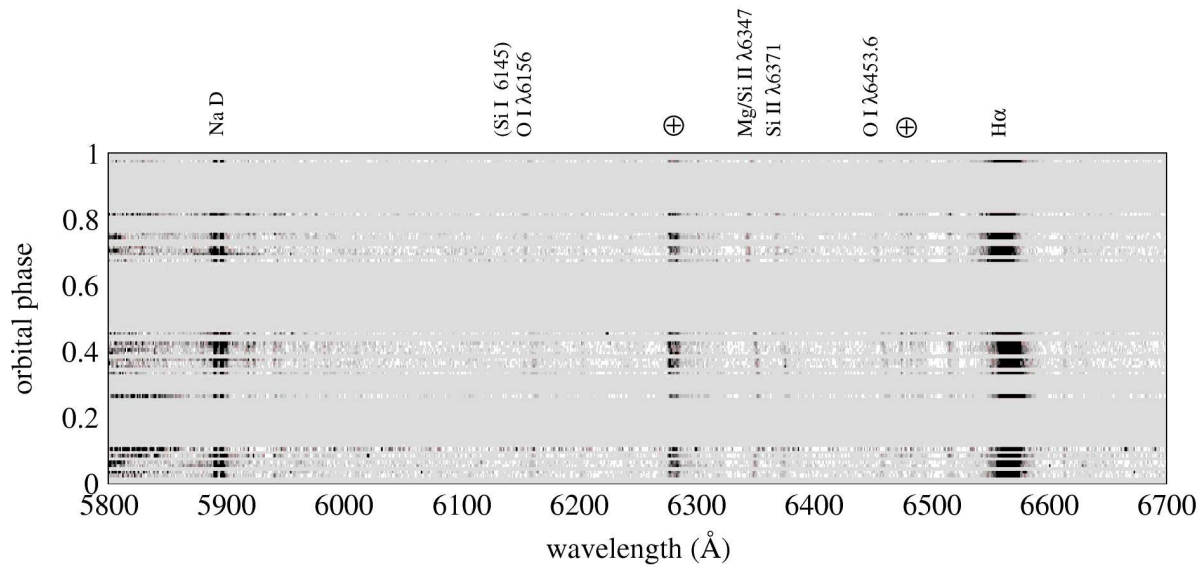}
\end{center}
\caption{Trailed quiescence spectra of V4641~Sgr.}
\label{trailed}
\end{figure*}

To check whether Eq.\ 1 is consistent with the spectroscopic
data, we also
phased the data with the spectroscopic ephemeris obtained by Orosz et al. (2001):
$P=2.81678$ d, $T_0=2451442.523$. The resulting phase diagram was shifted 
by $\Delta \phi=0.10\pm0.01$ in respect to fig.\ 2 of Orosz et al. (2001), i.e. the
minimum of the sine-wave occurred at $\phi=0.60\pm0.01$. We used this 
shift to correct the Orosz et al. period to $P=2.81720(2)$ d, indicating 
the good accuracy of the photometric period in Eq.\ 1.    

The best fit sine-wave in Fig.\ \ref{radvel} resulted in the following parameters:
$K_2=211.3\pm1.0$ km~s$^{-1}$, $\gamma=74.8\pm0.8$ km~s$^{-1}$. The $K_2$ velocity
amplitude is in excellent agreement with that of Orosz et al. (2001), who measured
$211.0\pm3.1$ km~s$^{-1}$. Combined with the period, the resulting mass-function is 
$f(M)=2.74\pm0.04$ M$_\odot$.  However, our $\gamma$-velocity differs by over 30
km~s$^{-1}$ (the Orosz et al. value is $\gamma=107.4\pm2.9$ km~s$^{-1}$), i.e. over
10$\sigma$.

To determine the cause of the discrepancy, we first checked the consistency between different radial velocity standards.
$\beta$~Vir, the other F-type standard, yielded essentially the same result, 
$\gamma=75.2\pm0.8$ km~s$^{-1}$, while HR~3383, the A1V
standard with broader H$\alpha$ line, gave $\gamma=68.0\pm0.7$ km~s$^{-1}$ (the $K_2$ 
radial velocity amplitude was the same in all cases within the error bars). Note that
the quoted errors came from the sine-wave fit, so that systematic errors are not
included. From these numbers, it is obvious, that {\it (i)} the large difference between 
the HD~187691 velocities and those of Orosz et al. (2001) is real and {\it (ii)} despite
the consistency of the two F-type standards, the HR~3383 velocities warn us that there
might be larger systematic errors due to spectral mismatch than the raw data would
suggest. In other words, it would be misleading to conclude from the two F-type standards
that we have measured the $\gamma$-velocity within $\pm0.6$ km~s$^{-1}$. On the other
hand, the broader H$\alpha$ line of HR~3383 may enhance possible systematics in 
cross-correlation velocities (for instance, due to uncertainties in continuum
normalization), so that the apparent offset between the F-type and
A-type standards ($\sim$7 km~s$^{-1}$) is likely to be more affected by the A-type 
HR~3383. Considering these aspects, we think the simple mean 
of the three values and its standard deviation give a more realistic $\gamma$-velocity 
and error estimate than, for instance, the average of the two F-type velocities.
We adopted therefore $\gamma=72.7\pm3.3$ km~s$^{-1}$ as our systemic velocity 
of V4641~Sgr.

We then noticed that in Sect.\ 2 of Orosz et al. (2001) its authors used a bright
night sky emission line to match the wavelength scales of spectra taken in different
observatories: {\it ``The bright night sky emission line at $\approx$5578 \AA\ was used
to make small adjustments to the wavelength of the FLWO and VLT spectra. The shifts
required to align this feature to a common wavelength (5578.0 \AA) were generally less
than $\approx$15 km~s$^{-1}$, although the VLT spectra from the end of the night of June
7 required shifts on the order of 80 km~s$^{-1}$.''} We checked this line and found that
it belongs to the neutral oxygen, for which the NIST Atomic Spectra 
Database\footnote{\tt http://physics.nist.gov/cgi-bin/AtData/lines\_form} lists 5577.339
\AA\ as the laboratory wavelength. On the other hand, IRAF's internal  telluric line list
({\tt linelists/skylines.dat}) contains two telluric features at 5577.3894 \AA\ and
5577.5948 \AA. These three wavelengths are $\sim22-36$ km~s$^{-1}$  towards blue from the
value used by Orosz et al. (2001), which suggests that the  Orosz et al. data are offset
by about 30 km~s$^{-1}$. If we adopt the O~I line from the NIST database, the Orosz et
al. data should be corrected by 35.5 km~s$^{-1}$, yielding a corrected $\gamma$-velocity
of $71.9\pm2.9$ km~s$^{-1}$, which is in perfect agreement with ours.  This is very
reassuring, as the $\sim30$ km~s$^{-1}$ shift in the centre-of-mass velocity over $\sim5$ years could otherwise be explained only by a very massive invisible third body. This not
only seems very unlikely, given the $14-20$ M$_\odot$ total mass of the
binary (Orosz et al. 2001), but also lacks support from other observations. 

Finally, we present the trailed quiescence spectra of V4641~Sgr 
in Fig.\ \ref{trailed}, in which we emphasize the weaker lines of the secondary star 
and the lack of spectral changes other than varying Doppler-shifts. Besides H$\alpha$, five further lines (of O, Mg and Si) clearly showed the orbital motion. Other lines remained constant (except for a $\sim$30 km~s$^{-1}$ scatter in the
telluric lines due to the heliocentric corrections calculated with the 
{\it rvcor} and {\it dopcor} tasks). The stability of these spectra indicated a minimum level of
mass-transfer activity during our observations.

\section{Discussion}

\subsection{Implications on outburst mechanisms}

In black-hole X-ray binaries, emission lines are generally from an accretion disk. They
have double-peak structures, typically with an equivalent  width of 10-20 \AA\ in
H$\alpha$ (see a recent discussion in Wu et al. 2002). Additionally, their line profiles often have weak
asymmetry due to orbital motion of a bright spot (e.g. Casares et al. 1995) or extended
structures around the accretion disk (Soria et al. 2000). 

The emission lines of V4641~Sgr are completely different from these tyical results.
Firstly, the lines are highly asymmetric, a feature which is difficult to attribute to
the weak contribution of a bright spot. Secondly, the equivalent width of the  H$\alpha$
is quite large, 130--170 \AA. The emitting area of lines should therefore have a volume
at least five times that of the entire disk which argues against scenarios with local
asymmetry in a disk. Furthermore, both Fe and He lines show P~Cygni profiles, which are
not expected from an accretion disk. Consequently, instead of emission from a disk, the
evidence points to a violent mass ejection and the presence of an outflow,  which began
optically thick in the Paschen lines and became optically thin within five hours. This
suggests that the ejection was very limited in time and our observations covered the
initial phase of shell expansion.

Although it has a very different character, we noticed a remarkable spectral
similarity with the X-ray transient CI~Cam (RXTE J0421+560). Robinson et al. (2002)
presented H$\alpha$ and Paschen line profiles of the 1998 March/April outburst of CI~Cam,
which had the same broad blue extensions like V4641~Sgr in outburst. Hynes et al. (2002),
whilst discussing their optical spectra which showed similar line profiles to ours, proposed a
two-component outflow for CI~Cam, with a fast, hot polar wind and a dense, cooler
equatorial outflow with much lower velocity. However, the temperature of the secondary
star in V4641~Sgr is too low to generate a strong stellar wind, so that the outflow is
presumably from a disk, in other words, a disk wind.

Regarding this point, we found another noteworthy similarity, this time between 
V4641~Sgr and the famous galactic jet source, SS~433. Spectra taken from it show an
H$\alpha$ emission line with both "moving" components, which are interpreted as optical
jets, and a "stationary" component (for examples of the latter, see Murdin et al. (1980),
Vermeulen et al. (1993) and Gies et al. (2002)). The profiles of this stationary line of
SS~433 depend on its orbital phase. Consequently, it is very interesting that the
emission line of V4641~Sgr, which was taken near an orbital phase of 0, is quite similar
to the stationary component of SS~433 when its orbital phase = 0. Given that the
stationary component in SS~433 is interpreted as originating from a disk wind, this
provides further evidence for the same in V4641~Sgr. As both systems have a high
inclination angle, the presence of a secondary star probably affects line profiles since
it would block a part of the wind at an orbital phase = 0. This may cause a weaker blue
component in H$\alpha$ and He lines of  V4641~Sgr.

Inspired by the similarity with SS~433, we unsuccessfully tried to find possible ``moving
emission lines''  for a sign of optical jets around the H$\alpha$ line of V4641~Sgr.
A possible feature can be seen at about 6360
\AA, but this is more likely to be due to atmospheric distortion of the Mg/Si~II
emissions. However, our spectral coverage was quite limited and we cannot firmly
exclude the existence of ``moving emission lines'' outside the 6047--7005 \AA\ region. 

From a theoretical point of view, disk winds have been studied for cataclysmic
variables, especially novalike systems. For example, Proga (2003b) calculated 
resonance-line profiles predicted by radiation-driven disk wind models for 
novalike systems. Some of his theoretical emission line profiles are similar to
lines in V4641~Sgr and SS~433. However, despite the recent progress in the 
field (e.g. Feldmeier \& Shlosman 1999, Feldmeier et al. 1999, Proga 2003a), 
the complexity of the problem prevents simple modeling of the observed line profiles. 
An important point is that this line-driven wind theory needs a strong UV radiation to
accelerate the wind. The strong UV radiation can be expected from the surface of white
dwarfs in cataclysmic variables. On the other hand, the UV radiation from black hole
X-ray binaries is expected to be weaker, especially in a low/hard state of X-ray
binaries. This is because a hot thermal accretion disk is truncated by an inner
``radiation inefficient accretion flow'' (RIAF).

The most fundamental difference between V4641~Sgr and SS~433 is their X-ray
luminosities. The X-ray luminosity of SS~433 reaches super-Eddington luminosity  for a
10 M$_\odot$ black hole ($\sim$10$^{39}$ ergs s$^{-1}$). According to theories of
super-Eddington accretion, the accreting gas forms an optically thick and geometrically
thin disk (so called ``slim disk'', Abramowicz et al. 1988).  This slim disk can
generate strong UV radiation, which is favourable for the line-driven wind model. But,
the X-ray luminosity of V4641~Sgr is about  $\sim$10$^{36-37}$ erg~s$^{-1}$ during 
minor outbursts . This value indicates a low/hard state, in which we cannot expect
strong UV radiation.

In summary, the driving mechanism of the disk wind in V4641 Sgr remains an open issue. It
is possible that the driving mechanism is not the UV radiation, but a magnetic field.
Another possibility is that the accretion rate actually reached the super-Eddington
level, but was undetected -- perhaps because most of the X-ray radiation was absorbed by
a thick torus around the black hole and reemitted in optical. Currently though, both scenarios
are merely speculations.

Another important question is: to what extent was the 1999 ``super-outburst'' different from
the smaller eruptions in 2002, 2003 and 2004? Based on the orders-of-magnitude lower
X-ray emission in the later outbursts, Uemura et al. (2004a, 2005) concluded that 
all minor outbursts had the same nature, which was different from that of the
1999 September event. Uemura et al. (2004a) also proposed a scenario for the time evolution
of the rapid brightness fluctuations: generation of a hot region in an
accretion disk, which propagates towards the inner portion of the disk, triggering
short-term fluctuations. When the active and bright region finally disappears, which is
observed as a dip, the whole cycle repeats with the replenishment of gas from the
outer region.

Our optical and near-infrared spectra suggest that despite the much weaker X-ray flux,
mass ejection had very similar kinematic properties in 1999 September and 2004 July. Most
noticeably, the description of the H$\alpha$ line in the 1999 outburst by Orosz et al.
(2001) and Chaty et al. (2003) fits our observations in 2004 perfectly. The broad, strong
emission lines and their variability (e.g. the He~I~$\lambda$5015 and  He~I~$\lambda$5049
lines) suggest the presence of a high-velocity outflow component blown off from the
accretion of matter on to the compact object. (Schulz \& Brandt 2002, Chaty et al. 2003).
The Paschen lines revealed the surprisingly short time-scale of this blow-off. The first
spectra contained strong emission in the upper lines, which is indicative of high
density/optically thick lines (e.g. Lynch et al. 2000). Within five hours, upper Paschen
lines strongly decreased, implying  switching to optically thin emission, so that the
blown-off matter did not have further supplies from the accretion disk. The variations of
the H$\alpha$ equivalent width during a 10-min flare-up revealed that rapid brightness
fluctuations are caused by continuum variations, therefore our observations are consistent
with the Uemura et al. (2004a) scenario in a sense that the source of these continuum
fluctuations must be in the inner part of the accretion disk. 

We also found subtle
changes in the H$\alpha$ line profile, which suggest the presence of additional phenomena during the
rapid fluctuations: whereas the redshifted emission component did not show any correlation
with the flare-up, the blue and the broad components exhibited significant changes in the
velocity range of the emitting gas. Moreover, the gradual shifts with opposite signs 
in the line centre of the blue and red components support the contention that the three components
(blue, red and broad) formed in well-separated parts of the system. The source of
the red component must have been located far out on the accretion disk, as it was left unchanged
by the flare. In contrast, the blue and broad components traced the
continuum changes very well, suggesting they must have originated much closer to the disk, with the broad
component being the closest.

\subsection{System parameters}

The outburst H$\alpha$ spectra allowed the detection of the 6613 \AA\ Diffuse
Interstellar Band (DIB) on the red wing of the H$\alpha$ profile. The equivalent width
of this DIB is known to correlate with the interstellar reddening; adopting
$EW_{6613}/E(B-V)\approx0.231$ (Jenniskens \& D\'esert 1994), the observed 
$EW_{6613}\approx0.04-0.06$ \AA\ yields $E(B-V)\approx0.22\pm0.05$ mag. The sodium D
doublet in quiescence is significantly stronger than reported for the 1999 outburst by 
Chaty et al. (2003): we measured the equivalent width of the sodium D$_1$ line  between
0.8 \AA--0.9 \AA\ in the best quiescence spectra, which is about twice as much  as the
0.45 \AA\ observed by Chaty et al. (2003). However, our spectra have reasonably better
resolving power ($R\sim6000$ vs. $\sim270$) and the given $EW_{D_1}$ suggests slightly
higher reddening, between 0.4--0.6 mag  (see fig.\ 4 in Munari \& Zwitter 1997).
Considering the large uncertainties, these reddenings are consistent with the previous
determinations:  $E(B-V)=0.32\pm0.10$ mag (Orosz et al. 2001) and $E(B-V)\approx0.25$
mag (Chaty et al. 2003).

Our radial velocity curve has the smallest uncertainties in the literature so far; the
$K_2$ velocity amplitude ($211.3\pm1.0$ km~s$^{-1}$) and the corresponding
mass-function  ($2.74\pm0.04$ M$_\odot$) are the same as those of by Orosz et al.
(2001).  The most significant improvement is found for the centre-of-mass velocity of
the system, for which we adopt the mean value of the three independent measurements:
$\gamma=72.7\pm3.3$ km~s$^{-1}$. Assuming that this systemic velocity is due
entirely to differential galactic rotation, we slightly revise the kinematical distance
limit to $d>6.3$ kpc, using the rotation curve given
in Fich et al. (1989) and the standard IAU rotation constants of $R_0=8.5$ kpc and
$\Theta_0=220$ km~s$^{-1}$ (from $d>7$ kpc, Orosz et al. 2001).  When combined with accurate
distance and proper motion measurements, the $\gamma$-velocity will be important
in finding the galactocentric orbit of V4641~Sgr. Knowing the galactic motion of an 
X-ray binary can help understand the possible origin of the system, which in turn can
reveal intriguing details of star formation in the early stages of evolution of the
Galaxy (Mirabel et al. 2001, 2002, Mirabel \& Rodrigues 2003). 

\section{Summary}

We have presented optical and near-infrared spectra of the galactic microquasar V4641~Sgr
in a short outburt in 2004 July and quiescence between 2004 September--2005 March. The
conclusions can be summarized as follows:

\begin{enumerate}

\item The photometric behaviour of the 2004 July outburst was similar to those in 2002
and 2003, i.e. the star showed rapid brightness fluctuations, due to changes in the
continuum radiation. However, the complex multi-component emission line profiles of 
hydrogen, helium and iron were similar to those observed in the big outburst  of 1999
September, with evidence of a similar high-velocity outflow and optically thick and 
then thin emission. This suggests that kinematic properties of the mass-ejection
processes were similar in 1999 and 2004 despite the huge difference in peak 
luminosities of the eruptions.

\item We have determined the most accurate radial velocity curve of the system  so far.
We confirm the optical mass function of the system by Orosz et al. (2001) reducing its
uncertainty by a factor of three to $f(M)=2.74\pm0.04$ M$_\odot$. We correct the systemic
velocity by more than 30 km~s$^{-1}$ to $\gamma=72.7\pm3.3$ km~s$^{-1}$. We do not find
spectral changes other than the Doppler-shifts due to the orbital motion,  so any
accretion activity must have been at low levels during our quiescence observations.

\item In the last couple of years it has been clearly established that V4641~Sgr has a 
recurrent activity with a time-scale of 1--2 years. Looking back, the shorter value 
seems to be more appropriate, which means every observing season should be monitored
for further eruptions. Our findings demonstrate the importance of taking rapid 
spectroscopic snapshots right after the discovery of
a new outburst. We suggest to keep the object on lists of possible target-of-opportunity
observations, because collecting more information on spectral changes in early phases 
is expected to yield better understanding of interactions around 
the black holes in systems like V4641~Sgr, especially the nature of mass ejection
and its connection to the accretion disk.

\end{enumerate}

\section*{Acknowledgments} 

This work has been supported by the OTKA Grant \#T042509 and the Australian Research
Council. C. Lindstr\o m received a Vacation Scholarship for this project from the School
of Physics,  University of Sydney. L.L. Kiss is supported by a University of Sydney
Postdoctoral Research Fellowship. A. Derekas joined the project within the International
Postgraduate Research Scholarship scheme of the Australian Department of Education,
Science and Training. The authors thank Prof. R.W. Hunstead for a careful  reading of the
manuscript. The NASA ADS Abstract Service was used to access data and references.

\end{document}